\documentclass[%
 reprint,
 amsmath,amssymb,
 aps,
]{revtex4-1}

\usepackage[utf8]{inputenc}
\usepackage{array}
\usepackage{amsmath}
\usepackage{slashed}
\usepackage{amsfonts}
\usepackage{blindtext}
\usepackage{newunicodechar}
\usepackage[pdftex]{graphicx}
\usepackage[all,cmtip]{xy}
\usepackage[english]{babel}
\usepackage{caption}
\usepackage{verbatim,upgreek}
\usepackage{array}
\usepackage{multirow}
\usepackage{amsfonts}
\usepackage{slashed} 
\usepackage{hyperref}
\usepackage{leading}
\usepackage{scalefnt}
\usepackage[dvipsnames]{xcolor}
\hypersetup{colorlinks=true,linkbordercolor=Blue,linkcolor=Blue, citecolor=Blue}
\usepackage{subcaption}
\usepackage{graphicx}
\usepackage{subeqnarray}
\usepackage{setspace}
\usepackage{indentfirst} %primeiro paragrafo com espaço
\usepackage{gensymb}%simbolo º
\newsavebox\myboxA
\newsavebox\myboxB
\newlength\mylenA
\begin{document}

%opening
\title{Search for Dark Sector by Repurposing the UVX Brazilian Synchrotron}

\author{L. Duarte,$^{1}$} 
\author{L. Lin,$^{2}$}
\author{M. Lindner,$^{3}$}
\author{V. Kozhuharov,$^{4}$}
\author{S. V. Kuleshov,$^{5,6}$}
\author{A. S. de Jesus,$^{1,7}$} 
\author{F. S. Queiroz,$^{1,5,7}$} \email{farinaldo.queiroz@ufrn.br}
\author{Y. Villamizar$^{1,7}$}
\author{H. Westfahl Jr, $^{2}$}

\affiliation{$^{1}$International Institute of Physics, Universidade Federal do Rio Grande do Norte,Campus Universit\'ario, Lagoa Nova, Natal-RN 59078-970, Brazil}
\affiliation{$^{2}$ Laborat\'orio Nacional de Luz Síncrotron - LNLS, Caixa Postal 6192, CEP 13084-971, CEP 13084-971 BRAZIL, Campinas} 
\affiliation{$^{3}$ Max Planck Institut fur Kernphysik, Saupfercheckweg 1, 69117 Heidelberg, Germany}
\affiliation{$^{4}$ Faculty of Physics, Sofia University, 5 J. Bourchier Blvd., 1164 Sofia, Bulgaria and INFN - LNF, Via E. Fermi 54 - 00044 Frascati, Italy}
\affiliation{$^{5}$ Millennium Institute for Subatomic Physics at High-Energy Frontier (SAPHIR), Fernandez Concha 700, Santiago, Chile} 
\affiliation{$^{6}$ Center for Theoretical and Experimental Particle Physics, Facultad de Ciencias Exactas, Universidad Andres Bello, Fernandez Concha 700, Santiago, Chile}
\affiliation{$^{7}$ Departamento de F\'isica, Universidade Federal do Rio Grande do Norte, 59078-970, Natal, RN, Brasil}

\begin{abstract}
We propose the first {\bf Se}arch for {\bf D}ark {\bf S}ector at the Brazilian Synchrotron Light Laboratory, site of Sirius, a fourth-generation storage ring. We show that UVX, Sirius predecessor, can be a promising dark sector detector, {\bf SeDS}, with unprecedented sensitivity.  The search is based on a $1-3$~GeV positron beam impinging on a thick target leading the $e^+ e^- \rightarrow \gamma A'$ reaction, followed by a missing mass spectrum event reconstruction. We show that {\bf SeDS} has the potential to probe dark photons with masses up to $55$~MeV and kinetic coupling down to $\epsilon  \sim  10^{-14}$ within months of data. Therefore, such experiment would constitute the best dark photon probe worldwide in the $10-55$~MeV mass range, being able to probe an unexplored region of parameter space.

\end{abstract}

\maketitle

\section{Introduction}
The Standard Model of particle physics has endured a multitude of precision tests over the past decades. The discovery of the Higgs Boson in 2012 constitutes a landmark \cite{ATLAS:2012yve,CMS:2012qbp}. Albeit, we are far from having a final theory that explains exciting observations, as dark matter. The LHC (Large Hadron Collider) has reached unprecedented energies, but yet has not been able to detect dark matter particles. The nature of dark matter is unknown, and its nature might be unveiled through the detection of a messenger that can be heavy or light, with no prejudice. As no positive signal has been observed from WIMPs (Weakly Interacting Massive Particles), there is a growing interest in the community for light dark matte particles that belong to a
dark sector, which may feature light mediators. Such light force carriers can have different interactions with SM particles and are subject to a multitude of experimental searches by low energy accelerators \cite{Alexander:2016aln}. Among several possibilities, a vector mediator usually called dark photon or  hidden photon has been subject to a multitude of studies. The dark photon is the simplest interaction that can be tested experimentally because it represents a dark Quantum Electrodynamics. The dark photon interacts with the photon through the gauge-invariant lagrangian \cite{Holdom:1985ag,Fayet:1990wx},

\begin{equation}
\mathcal{L} = - \frac{\epsilon}{2} F^{\mu \nu} F^{\prime \mu \nu}.
\end{equation}

The presence of such a term is universal to all Abelian extensions of the SM,  regardless if the new gauge symmetry is broken at small or high-energy scales. As a result, the dark photon ($A^\prime$) inherits an electromagnetic interaction proportional to the kinetic mixing term, $\epsilon$, which is governed by \cite{Pospelov:2007mp,Fayet:2007ua},
\begin{eqnarray}\label{lagrangian}
 \mathcal{L}\supset -e\epsilon J^{\mu}A'_{\mu},
 \label{eq1}
\end{eqnarray}
where $J^{\mu}$ is the electromagnetic current, $e$ the electric charge. Notice that $\epsilon$ links the dark and visible sectors, i.e. the SM spectrum. The mass of the dark photon can be treated as a free parameter \cite{Fabbrichesi:2020wbt}. Without loss of generality, we will consider the case in which $A^\prime$ couples to leptons. 

\begin{figure}[!t]
    \centering
    \includegraphics[width=\columnwidth]{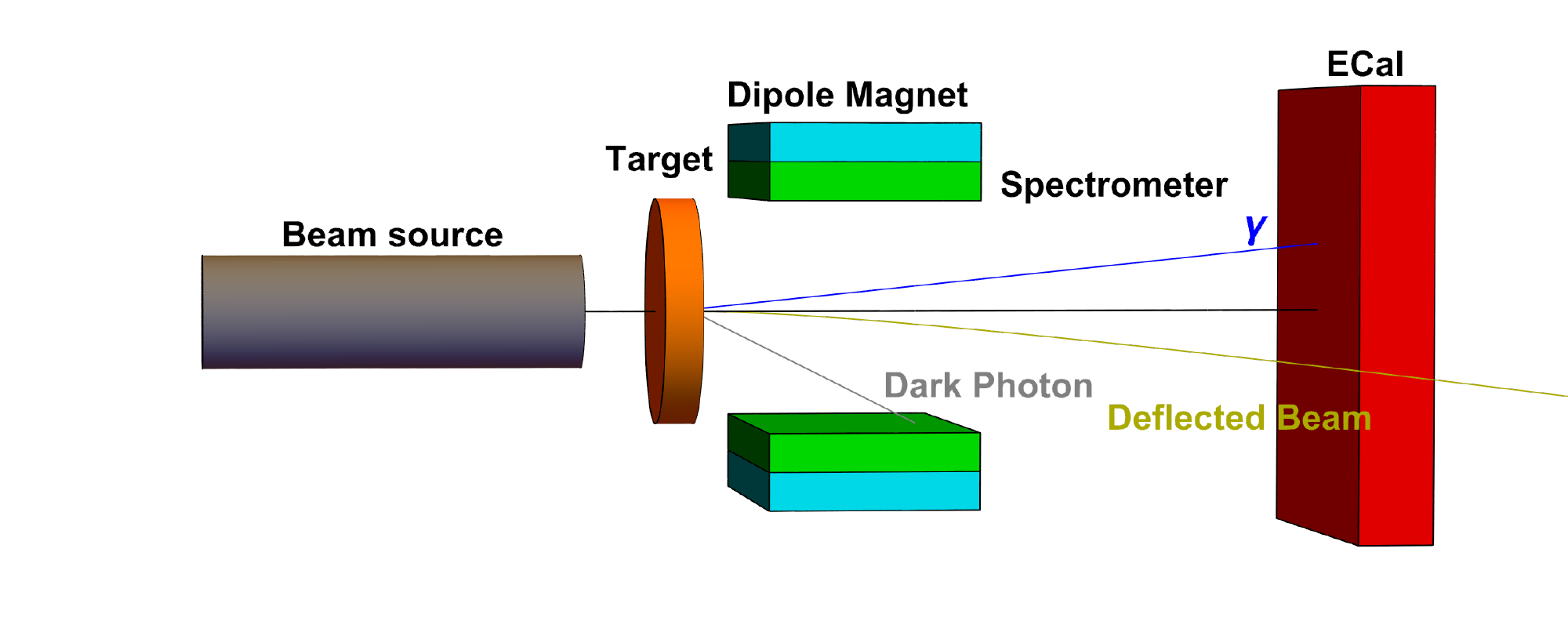}
    \caption{A schematic illustration of {\bf SeDS} experiment devoted to search for Dark Sectors.}
    \label{experimentdescription}
\end{figure}

The search for a new force carrier has motivated many theoretical and phenomenological studies \cite{deRomeri:2020kno, Breitbach:2021gvv, Batell:2009yf,Graham:2021ggy}, stimulated the reanalysis and interpretation of old data \cite{deNiverville:2016rqh, BaBar:2017tiz}, and promoted new experimental programs \cite{Raggi:2014zpa, Rachek:2017gdc, alexander2017mmaps, SHiP:2018yqc}. Our work presents a new proposal devoted to the search for dark photons at the Brazilian Synchrotron Light Laboratory (lnls), a new positron on target experiment with features that makes it a promising dark sector detector.

\section{SeDS }
The second-generation Brazilian synchrotron light source, UVX, was a $1.37$~GeV electron storage ring recently decommissioned in Campinas, Brazil. Its injection system included a $120$~MeV linear accelerator and a $500$~MeV Booster Synchrotron Injector \cite{Lin:1993np,Liu:2010bfa,Lescano:2017vlq}. UVX has now been succeeded by Sirius, a fourth-generation store ring \cite{Rodrigues:2019oej,Alves:2019aei,Liu:2021txs}. Several subsystems of UVX could potentially be used to repurpose this old light source into a new $1-3$~GeV positron accelerator to host a new SeDS small-scale fixed target experiment set to look for dark photons via the process $e^+ e^- \to \gamma A'$. The mass of the dark photon can be determined using the missing mass technique, which requires the knowledge of the initial parameters of the positron beam, the electron target at rest, and the energy-momentum of the final state photon. Hence, the only assumption of the proposed experimental technique is that the dark photon couples to leptons.
From a realistic perspective, we expect to be competitive in the search for dark photons through the annihilation process, since the positron beam at {\bf SeDS} can reach $1$~GeV, and optimistically $3$~GeV, which is not achieved by current experiments working with the same technique \cite{PADME:2021vjp, Rachek:2017gdc}.

\section{Experimental strategy and kinematics}

The experimental design of the experiment is exhibited in Fig.\ref{experimentdescription},  where accelerated positrons are directed to a diamond target, producing photons and $A^\prime$. 
We highlight that a Carbon target could be used instead, but we selected diamond mostly because it acts as a solid-state ionization chamber, and has very good thermal conductivity, which is appropriate for high beam intensity, as one can dissipate heat through the side connections. The photons are expected to hit an electromagnetic calorimeter to extract the properties of the final state, and we have a spectrometer to measure charged interactions in a momentum range. The dipole magnet is added to deflect the positron beam and reduce the number of events on the calorimeter. 
% OK, it is fine. Actually the thermal conductivity is the important feature. For PADME we calculated that
% the increase of temperature, in the case of no cooling and heat dissipation, was ~ 1 deg. C per day.
%The dominant process of energy loss of the positrons in the target is the radiation (since
%$\beta\gamma$ > 1000 and the critical energy for carbon is about 85 MeV). However, majority of the radiated
%bremsstrahlung photons leave the target and the thermal load on the target is dominated by
%the ionization energy losses of the beam positrons. 

\begin{figure}[!h]
\centering
%\subfloat[Fluxo de neutrinos.]
\hspace{-0.0cm}{
\includegraphics[scale=0.35]{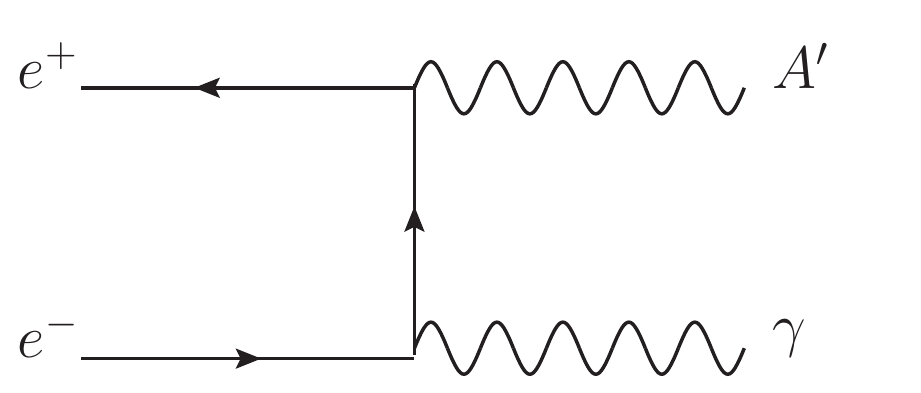}
\includegraphics[scale=0.35]{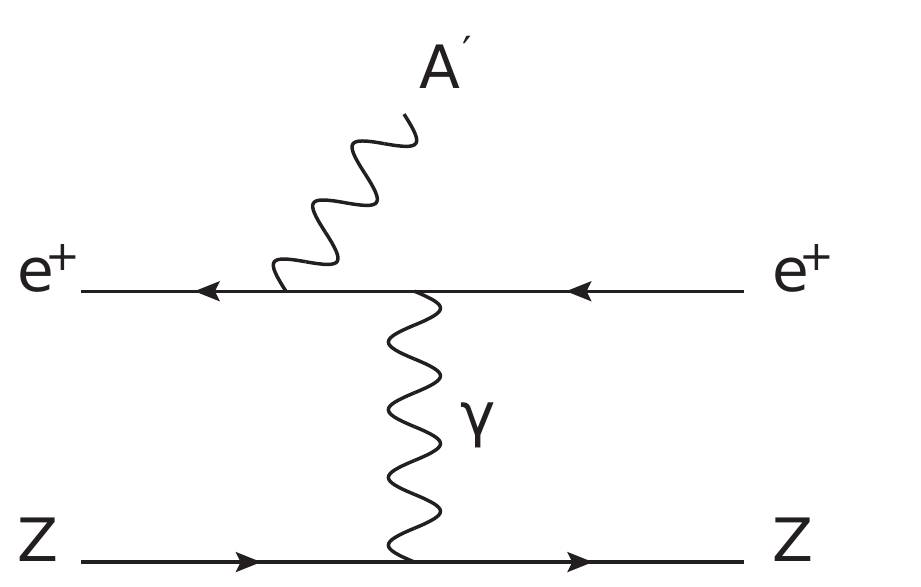}
}
\caption{Dark photon production mechanisms by high energy positrons on a fixed target. Left: $A'$-production in $e^+e^-$ annihilation. Right: $A'$-production via Bremsstrahlung.}
\label{feynmanDiagrams}
\end{figure}

The processes involved in dark photon production by
$(\mathcal{O})$\,GeV positrons impinging on a thin target are $e^+ e^- \to A'\gamma$ and $e^+ Z \to e^+ Z A'$, the so-called annihilation and $A'$-strahlung production.
Annihilation processes occur when dark photons interact with electrons in the material of the target, shield, and
detector, thereby producing real photons in a process similar to Compton scattering. 
The $A'$-strahlung, instead, represents the radiative $A'$ emission by an impinging $e^+$ in the electromagnetic field of a target nucleus.
Both processes are similar to the ones for ordinary photons, as shown in Fig. \ref{feynmanDiagrams}, and their cross-section scale with $\epsilon^2$. 

The main goal of the {\bf SeDS} experiment is to search for a dark photon produced in the annihilation
of positrons of the beam and electrons at rest in a diamond target. This material has a low atomic number ($Z=6$) which allows limiting the bremsstrahlung interactions (cross-section proportional to $Z^2$), which is the main background in annihilation searches \cite{Simeonov:2021oin, Ciafaloni:2020cic}. Choosing a graphite target (which has the same atomic number $Z = 6$), the experiment would have to run for a much longer period to access the same sensitivity found with the diamond target because of its lower density. 
Furthermore, a diamond target produced by chemical vapor deposition process is not much more expensive.
A dark photon signal is assed by missing-mass events. The ordinary photon in the final state can be observed and its four-momentum measured. The invisble 
$A^\prime$ will appear as a bump in the missing mass spectrum,

\begin{eqnarray}
 M^2_{\text{miss}}=(p_{e^-} + p_{\text{beam}}-p_{\gamma})^2,
\end{eqnarray}
and no assumptions about the $A'$ decay mode are indeed necessary.
Two-photon annihilation is the dominant process of high-energy photon production.
The emission angle of the final photon $\theta_{\gamma}$ with respect to the direction of the positron beam defines the value of the photon energy $E_{\gamma}$.  In the case of
two-photon production $E_{\gamma\gamma}^{\text{lab}} \approx E_{\text{beam}}(1-\cos\theta_{\gamma}^{\text{CM}})$, whereas for $A'$-boson production: $E_{A'\gamma}=E_{\gamma\gamma}^{\text{lab}}(1-M_{A'}^2/s)$. The maximum positron energy of $1$ and $3$\,GeV allows the production of $A'$ bosons through annihilation up to a center of mass (CM) energy $\sqrt{s}=\sqrt{2E_{\text {beam}} m_e}=31.9$ and $55.3$~MeV, respectively, where $m_e$ is the electron mass. The experiment luminosity can be computed using the relation,
\begin{eqnarray}\label{Lumi}
 L_{\text{inst}} = \frac{\text{P.O.T}}{\text{s}} N_{A} \frac{Z\rho d}{A}. 
\end{eqnarray}
{\bf SeDS} would be a fixed target experiment with a diamond target of $d=100-500\mu$m. 
Its instantaneous and integrated luminosities can be calculated using $Z=6$, density of diamond, $\rho=3,51$\,g/cm$^{3}$, and $A = 12.01$\,g diamond's gram-molecular weight. The experiment can generate $10$ bunches per second with $10^9$ positrons in each bunch, which corresponds to $10^{10}$ positrons on target (P.O.T) per second. 
%
%The parameters are summarized in Table \ref{tab:1}.
%\begin{table}[!htb]
%\scalefont{1.3}
%\centering
%\begin{tabular}{c c c}
%\hline
% Beam & & $e^+$  \\ 
%\hline
%Maximal Beam energy [GeV]& & $3$\,GeV \\
%Beam rate [particles/bunch]& & $10^9$ \\
%Number of bunches per second [Hz]& & $10$ \\
%Max. average current during a bunch [mA]& & $10$ \\
%Bunch lenght [ns] & & $10$\\
%\hline
%\end{tabular}
%\caption{SIRIUS beam parameters.}
%\label{tab:1}
%\end{table}
%
%Considering one year of taking data correspond to $10^7$\,s of effective experiment run time, the integrated luminosity will be 
%\begin{eqnarray}
%\mathcal{L}_{\text{int}}\sim 10^3 \text{pb}^{-1}/\text{Year}.
%\end{eqnarray}
%
\section{Background}
Several processes can be identified as potential background sources.
The cross-sections for the major background contributions at
$E_{\text{beam}} = 1~\text{GeV}$ are shown in table \ref{tab:bkg}.
The value of 30 MeV cut on the gamma energy was chosen as 1\% 
of the maximal beam energy, assuming a calorimeter is able to achieve
such resolution on the energy measurement.

\begin{table}
\centering
\begin{tabular}{l|c|c} \hline
Process & $\sigma$ @ 1 GeV [mb]& $\sigma$ @ 3 GeV [mb]  \\
\hline
$e^+e^-\to\gamma\gamma$ & 0.93 &  0.36  \\
$e^+Z \to e^+ Z \gamma$ & $2.2\times 10^3$ & $2.9\times10^3$\\
$e^+e^-\to\gamma\gamma\gamma$ & 0.02 & 0.016\\
$e^+e^-\to e^+e^- \gamma$  & 77  &  135\\
\hline
\end{tabular}
\caption{Cross-sections of the dominant background 
contributions to the search for $e^+ e^-\to\gamma A'$. 
$E_{\gamma} > 30~\text{MeV}$ for the infrared divergent processes.}
\label{tab:bkg}
\end{table}

\begin{figure}[!h]
\centering
\includegraphics[width=0.8\columnwidth]{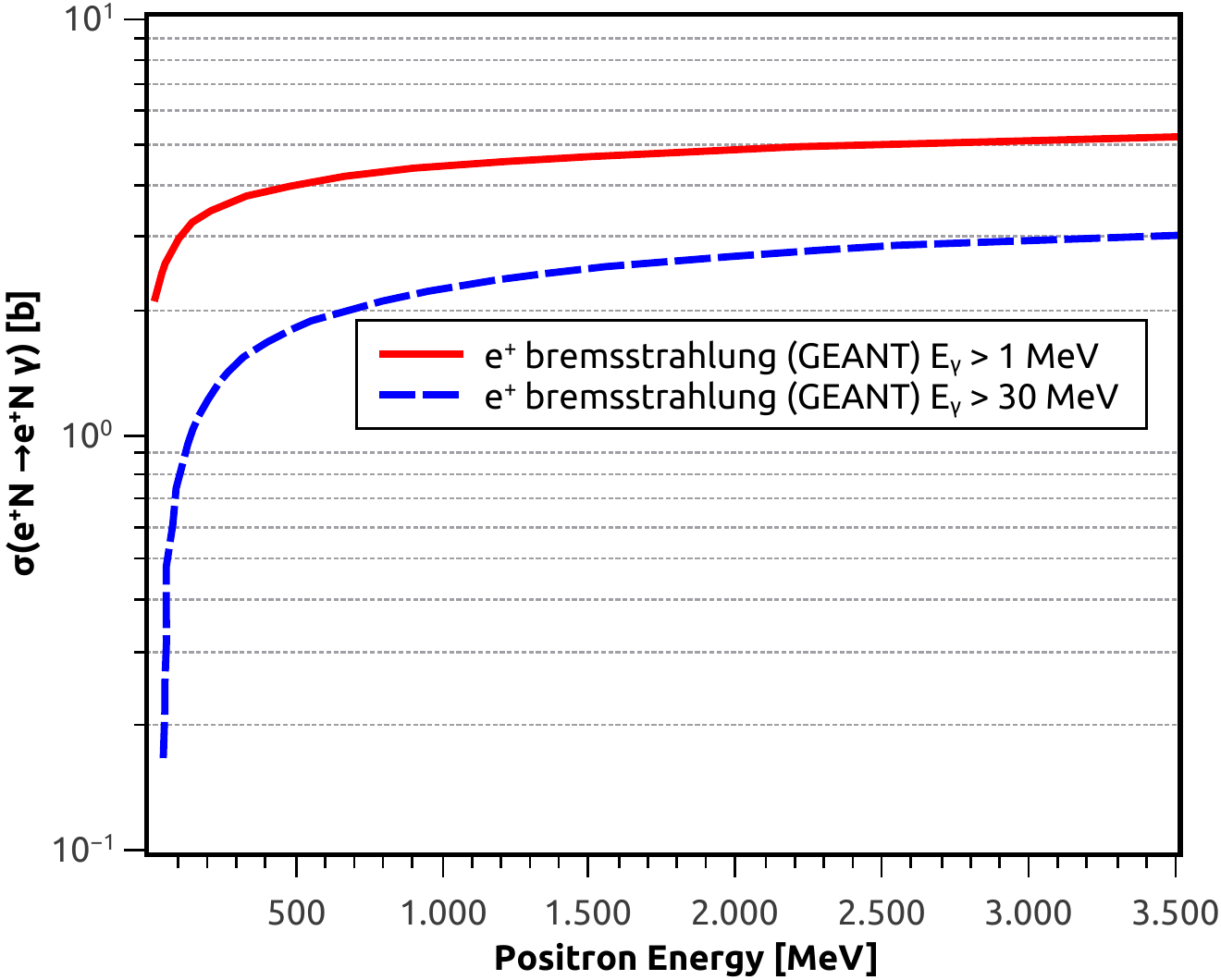}
\caption{Cross-section for emission of a high energy photon for as a function 
of the beam energy.}
\label{fig:brems}
\end{figure}
The dominant contribution is the radiation of a 
hard photon $\gamma$
from the beam positrons, either due to 
bremsstrahlung in the field of the {diamond} nuclei $Z$, 
or due to radiative Bhabha scattering. 
The bremsstrahlung cross-section increases 
logarithmically with the beam energy (fig. \ref{fig:brems}, \cite{GEANT4:2002zbu}),  but the opening angle $\langle\theta_{\gamma}\rangle$ 
of the emitted photon decreases as $1/E$ 
and the emission is highly forward.  
For the $e^+e^- \to \gamma A'$ process, the invariant mass scales as $\sqrt{E}$, and
the opening angle $\langle\theta_{\gamma}\rangle$ for the recoil photon  shrinks with  $1/\sqrt{E}$.
% OK, approved

The two photon annihilation can be kept under control due 
to the symmetric kinematics of the outgoing photons. 
%In addition, the successful reconstruction of 
%$e^+e^- \to \gamma\gamma$ will provided 
%the necessary normalization. 
The three photon annihilation, despite the small cross-section, 
is one of the dominant backgrounds due to the loss of 
symmetry in the event and the non-peaking 
distribution of $M_{miss}^2$. 
The employed photon vetoing system should be based 
on detectors with high efficiency and hermeticity.

The background suppression will govern the design of the detector system. A high granularity electromagnetic calorimeter will be necessary to precisely measure the $\theta_{\gamma}$
% OK, approved
of the impinging photons and provide sufficient separation and identification of
possibly overlapping showers.
A full simulation of the experimental setup is necessary to reliably assess the background contribution, 
together with a simulation of the major characteristics of the 
chosen readout system. Since this is outside the scope of our work, 
an indication of the physics potential of the proposed 
experiment is presented by the single event sensitivity, under the assumption that we can identify and/or veto the background events.

\section{Projected sensitivity for $A'$ production at lnls}\label{sec:DFSIRIUS}

In the experimental setup proposed, the dark photon is produced with a total cross-section at tree-level that reads, 

\begin{eqnarray}
\sigma_{\text{ann.}}=\frac{8\pi\alpha^2\epsilon^2}{s}\left[\left(\frac{s-m_{A'}^2}{2s} + \frac{m_{A'}^2}{s-m_{A'}^2}\right)\log\frac{s}{m_e^2} \right. \nonumber \\
\left.- \frac{s-m_{A'}^2}{2s}\right]. 
\label{eqCS}
\end{eqnarray}

\begin{figure}[!h]
\centering
\includegraphics[width=\columnwidth]{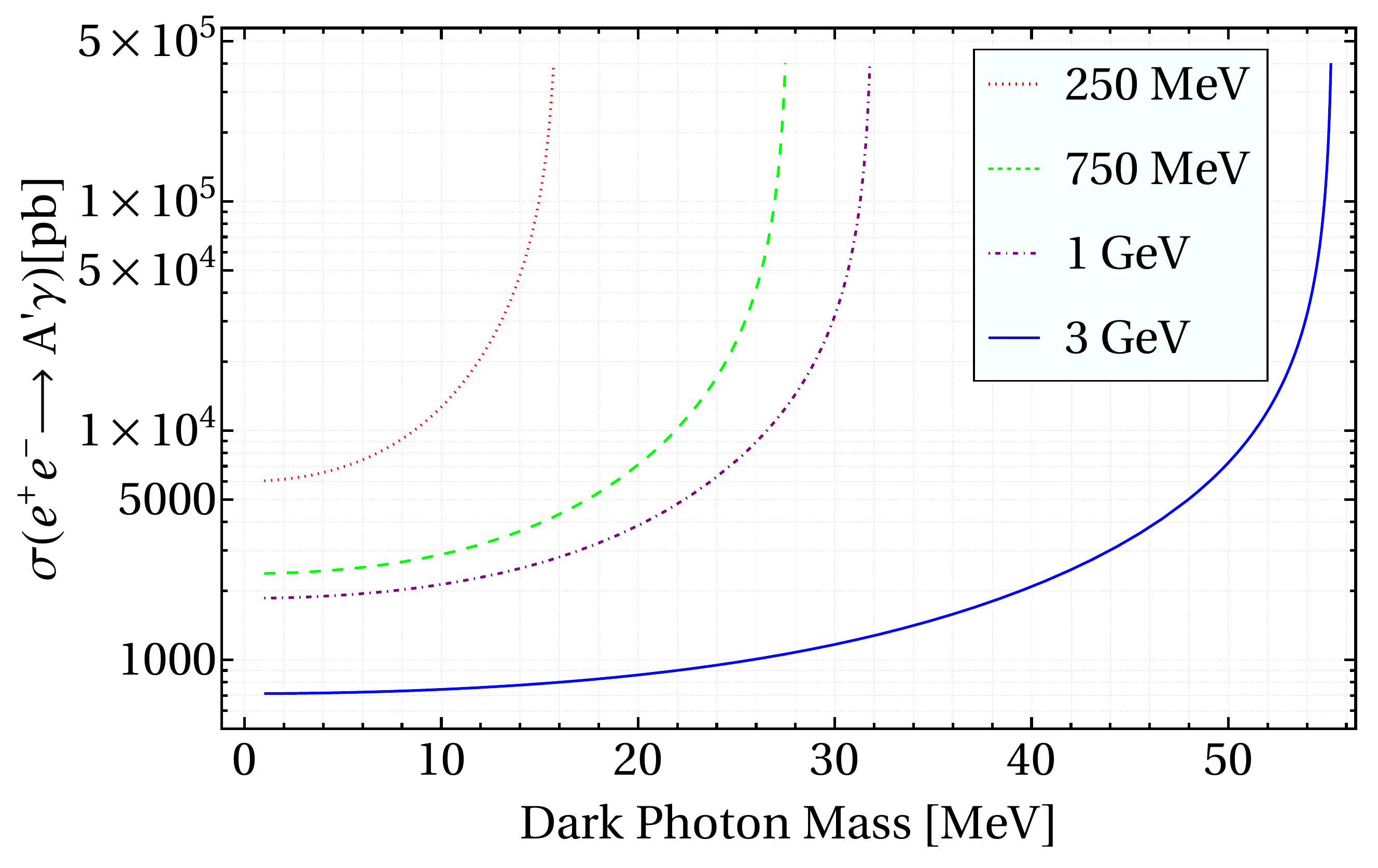}
\caption{$A'$ boson production cross-section as a function of its mass for different beam energies, $E=250$~MeV,$750$~MeV and $3$~GeV. We adopted $\epsilon=10^{-3}$.}
\label{cs}
\end{figure}

The expected annihilation cross-section as a function of the mass of the dark photon for different beam energies is displayed in Fig. \ref{cs}. Notice that when we increase the energy beam, the cross-section decreases, and this feature translates into a smaller sensitivity on the kinetic mixing for a given dark photon mass. On the other hand, the larger the beam energy, the larger the dark photon mass kinematically accessible in this process.

\begin{figure*}[t]
\centering
\includegraphics[width=0.92\columnwidth]{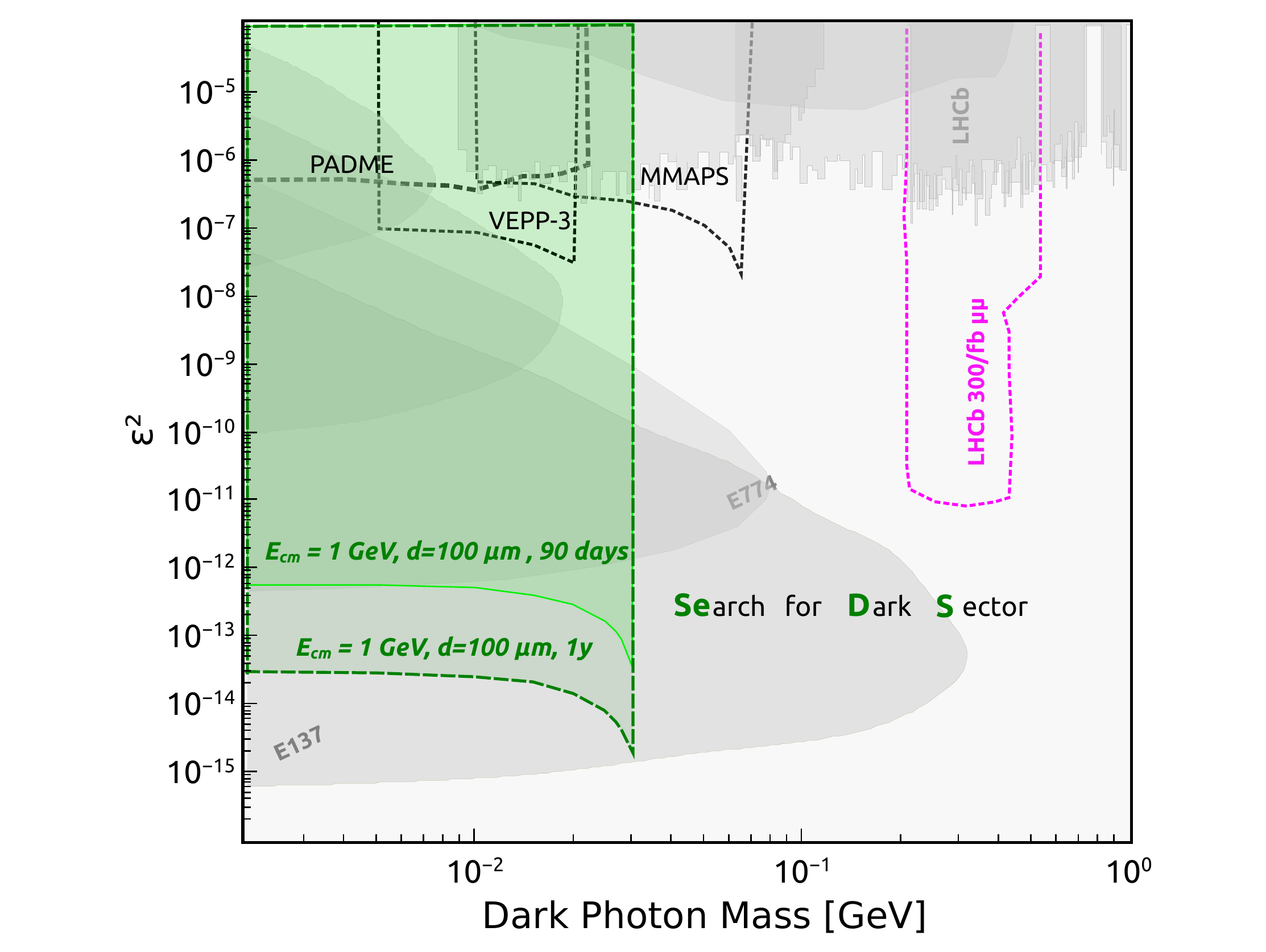}
\includegraphics[width=1\columnwidth]{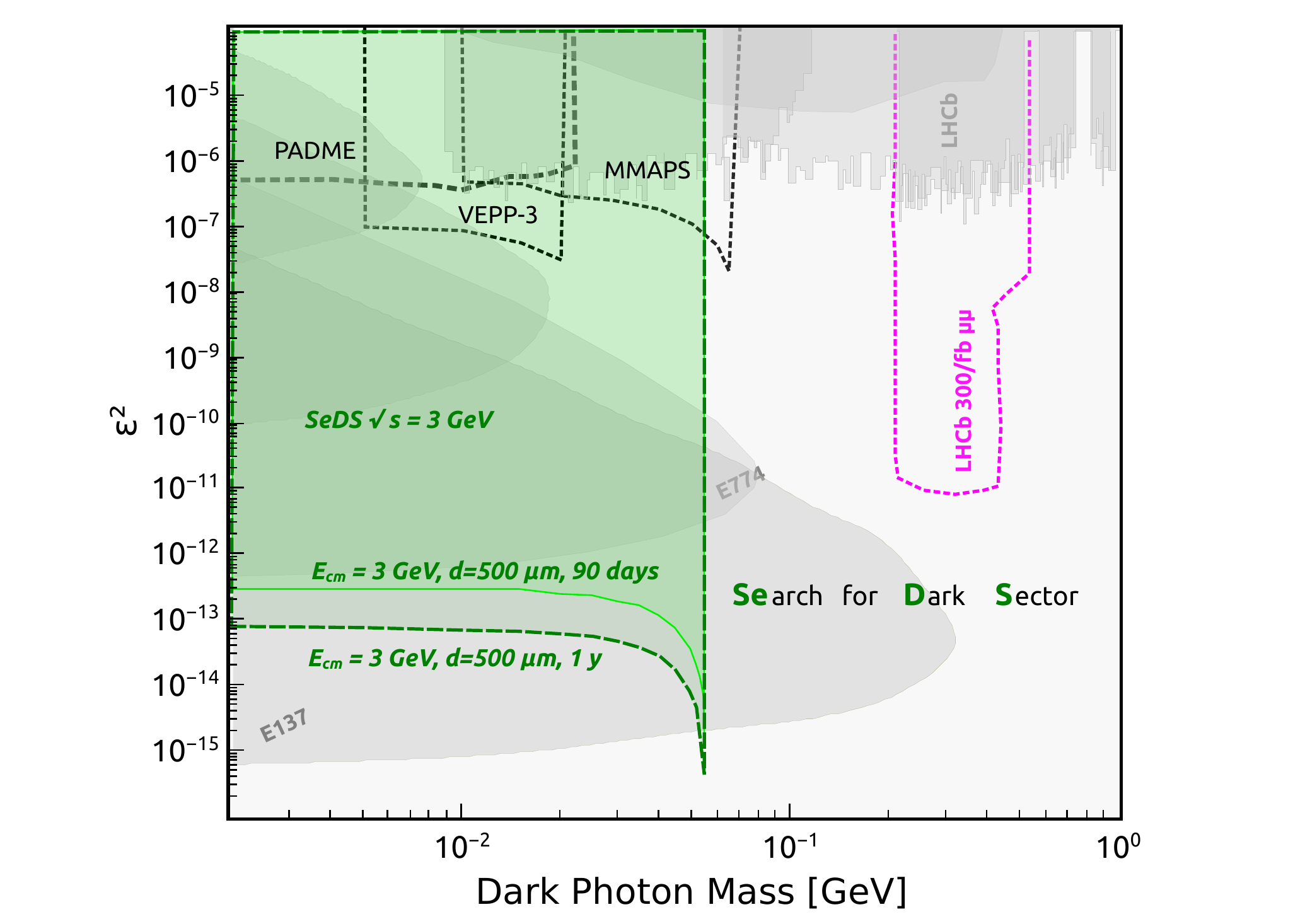}
\caption{In the left-panel we display {\bf SeDS} sensitivity on $\varepsilon^2$ as a function of dark photon mass for a 1GeV positron beam impinging on a diamond target of $100\mu$m  with 90 days and one-year of data. In the right-panel, we adopted a 3GeV positron beam with a diamond target of $500\mu$m, instead. }
\label{visible}
\end{figure*}

Assuming that we can control the background events as discussed in the previous sections, the statistical significance can be computed in the large sample limit, under the assumption that the signal events, after the cuts, are much larger than the background events  \cite{Elwood:2018qsr}. With this procedure, we find Fig.\ref{visible}, which displays the projected exclusion regions in the $\{M_{A'}, \varepsilon^2\}$ plane at 68\% C.L. In the left-panel we plot the projected sensitivity of {\bf SeDS} (green shaded regions) operating with a positron beam of $1$~GeV with a diamond target of $100\mu$m, for 90 days ($\mathcal{L}= 825$~pb), and one year ($\mathcal{L}= 3300$~pb). A portion of the parameter space has been excluded by previous experiments represented by gray contours, but there is a region for $\epsilon^2$ between $10^{-6}- 10^{-9}$ and $M_{A^\prime} \sim$ 10 MeV-30 MeV that {\bf SeDS} could potentially discover dark photons. It is clear that when we increase the luminosity, the projected exclusion region improves and that reflects in a shift towards lower values of $\epsilon$. The sharp increase in sensitivity for $M_{A^\prime} =30$~MeV is the result of the denominator in Eq.\ref{eqCS} that has a $s-m_{A'}^2$ term, leading to a large cross-section when the dark photon mass approaches the CM. A similar feature occurs in the right-panel, where we repeat our exercise for a positron beam of $3$~GeV, but there we adopt a diamond target of $500 \mu$m, which boosts our luminosity by a factor of five see (Eq.\ref{Lumi}). As aforementioned the beam energy dictates the largest dark photon mass probed in the annihilation process. Hence, with a $3$~GeV positron beam, $\sqrt{s}=55.3$~MeV, the sensitivity region towards larger dark photon masses. However, as we increase the CM, the production cross-section decreases, see Eq.\ref{eqCS}, weakening our sensitivity on $\epsilon$, except for the resonance peak. Notoriously, with a $3$~GeV positron beam, we can probe a sizeable unexplored region of parameter space with the potential to discover the presence of a dark photon, i.e. a new force carrier in nature. 
%{\bf Even if we are unable to use all positrons on target generated,  which would result in a smaller luminosity, and consequently shift our sensitivity to larger values of $\epsilon$, it is clear from Fig.\ref{visible}, we would still probe a large unexplored region of parameter}. 
Even if we are unable to utilize all available positrons on target,  
which would result in a smaller luminosity, 
and consequently shift our sensitivity to larger values of $\epsilon$, 
it is clear from Fig.\ref{visible}, we would still probe a large unexplored region of parameter,
currently inaccessible by other techniques. 
%Some small editing (VK), otherwise it's OK

In the figures, we overlay the expected exclusion bounds from planned experiments based on the missing mass technique, namely \textit{PADME} which uses a $550$~MeV positron beam, and consequently can probe a dark photon up to masses of $23.7$~MeV and $\epsilon^2\sim 10^{-8}$ \cite{PADME:2020ljb}; \textit{VEPP3} which aims to hit a positron beam of 500MeV on a gas of hydrogen, that could probe $\epsilon^2=3\times10^{-8}$ for $M_{A^\prime}=5-20$~MeV \cite{Rachek:2017gdc, wojtsekhowski2018searching}; \textit{MMAPS} that features a much more energetic positron beam of $6$~GeV incident on thick beryllium target, aimed to reach $\epsilon \sim 10^{-8}$ for $M_{A^\prime}=20-78$~MeV \cite{alexander2017mmaps}.  Unfortunately, PADME is the only experiment currently taking data \cite{Domenici:2020poy}, and as shown in Fig.\ref{visible} the projected limit falls into a region that could be fully covered by {\bf SeDS}. 
\section{Conclusions}
We have proposed the first search for dark sector at the Brazilian Light Source Laboratory using UVX, a second-generation store ring, which is currently decommissioned and could potentially be repurposed to search for dark photons, reaching unprecedented sensitivity. It has the potential to probe an unexplored region of parameter space, for $\epsilon \sim 10^{-6} -10^{-9}$ and $M_{A^\prime}=10-30$~MeV within 90 live-days using $1$~GeV positron impinging on a diamond target of $100\mu$m. We have shown that this sensitivity could be greatly improved by either using a thicker target or, more costly, increasing the energy of the positron beam. 

\acknowledgments
We thank Paolo Crivelli, Claudio Dib, Alfonso Zerwekh, and Sergey Kovalenko for discussions. This work was financially supported by Simons Foundation (Award Number:884966, AF), FAPESP grant 2021/01089-1, ICTP-SAIFR FAPESP grant 2016/01343-7, CAPES under Grant No. 88882.375870/2019-01, CNPq grant 408295/2021-0, Serrapilheira Foundation (grant number Serra-1912–31613), FONDECYT Grant 1191103 (Chile)  and ANID-Programa Milenio-code ICN2019\_044. 

\bibliographystyle{unsrt}
\bibliography{references}

\end{document}